# Atomic-scale imaging of electronic nematicity in ferropnictides

Qiang-Jun Cheng[1], Yong-Wei Wang[1], Ming-Qiang Ren[1,2], Ze-Xian Deng[1], Cong-Cong Lou[1],

Xu-Cun Ma[1,3*], Qi-Kun Xue[1,2,3,4*], Can-Li Song[1,3*]

[1]*State Key Laboratory of Low-Dimensional Quantum Physics, Department of Physics, Tsinghua University, Beijing 100084, China*

[2]*Shenzhen Institute for Quantum Science and Engineering and Department of Physics, Southern University of Science and Technology, Shenzhen 518055, China*

[3]*Frontier Science Center for Quantum Information, Beijing 100084, China*

[4]*Beijing Academy of Quantum Information Sciences, Beijing 100193, China*

Electronic nematicity, a correlated state characterized by broken rotational symmetry, has been recognized as a ubiquitous feature intertwined with unconventional electron pairing in various iron-based superconductors. Here we employ spectroscopic-imaging scanning tunneling microscopy to visualize atomic-scale electronic nematicity directly on FeAs planes of a prototypical ferropnictide $BaFe_2As_2$. Spatially, the nematic order appears as $4a_{Fe}$-spaced stripes ($a_{Fe} \sim 2.8$ Å is the in-plane Fe-Fe distance) within homogeneously and orthogonally oriented nano-domains. The energy-resolved conductance maps reveal a pronounced energy-dependence of the nematic order parameter that experiences a sign change at approximately 30 meV. This characteristic behavior coincides with energy-dependent orbital splitting previously identified in momentum space, but is remarkably visualized in real space for the first time in our study. Moreover, the electronic nematicity exhibits pronounced sensitivity to single impurities and is notably suppressed by cobalt substitution for Fe atoms, promoting optimal superconductivity when nematic fluctuations are strongest. Our results provide pivotal experimental insights for developing a microscopic model of nematic order, thus paving the way to study its complex relationship with unconventional superconductivity.

*Correspondence to: clsong07@mail.tsinghua.edu.cn, xucunma@mail.tsinghua.edu.cn, qkxue@mail.tsinghua.edu.cn

**Introduction**

Electronic nematicity, a state of matter that spontaneously breaks rotational symmetry while preserving translational symmetry, was first brought into prominence in iron-based superconductors (IBSCs)[1-9]. Although this exotic state has since been observed in an ever-growing range of correlated materials, such as cuprates[10-12], oxide films[13,14], kagome superconductors[15-17] and twisted graphene[18,19], its most thoroughly studied and crucial manifestations remain in IBSCs. Experimentally, nematic states in IBSCs have been identified through a huge variety of probes, including the tetragonal-to-orthorhombic structural phase transition[20,21], energy splitting of the degenerate $3d_{xz}$ and $3d_{yz}$ orbitals of Fe atoms[22-24], in-plane anisotropies in resistivity[25] and spin susceptibility[26-28]. Based on these measurements, nematic order and/or fluctuations were found to extend across a broad regime of the phase diagram, closely proximate to and intricately intertwined with superconductivity in IBSCs[29]. This has sparked significant interest in uncovering the mystery of electronic nematicity, as it might provide valuable insights into the nature of unconventional high-temperature ($T_c$) superconductivity. However, the intimate coupling between lattice, spin and charge degrees of freedom makes it challenging to identify the key driving force of nematicity in IBSCs[30-32]. A substantial body of work has revealed the characteristic electronic nematicity emerging in regimes with anisotropic spin fluctuations and/or orbital order. In theory, stripe-like magnetic correlations naturally induce an asymmetry between the $x$ and $y$ directions (collinear antiferromagnetic order in IBSCs), which in turn gives rise to anisotropic electron dispersion, manifested as orbital order[30].

Among all experimental measurements, direct imaging of the nematic order parameter is highly desired to advance our understanding of the underlying mechanism of electronic nematicity. In this context, angle-resolved photoemission spectroscopy (ARPES) has played a crucial role in detecting the universal momentum dependence of the energy splitting of Fe $3d_{xz/yz}$ orbitals as reported in FeSe and a prototypical ferropnictide compound $BaFe_2As_2$[23]. Subsequently, the nematicity has been also macroscopically detected via laser photoemission electron microscope, revealing mesoscopic nematic waves in IBSCs[33]. To directly visualize nematic patterns in real space, spectroscopic-imaging scanning tunneling microscopy (abbreviated as SI-STM), with its unrivaled spatial and energy resolutions, has emerged as an indispensable tool for probing the microscopic-scale properties of electronic nematicity. For example, maze-like or strain-stabilized domains have been observed in stoichiometric FeSe[34] and LiFeAs[35], yet within each domain, unidirectional charge modulations with varying spatial periods have been unexpectedly visualized, leading to the breaking of both rotational and translational symmetries. In contrast, nanoscale nematicity was reported in NaFeAs with novel energy and temperature dependence, highlighting the presence of pronounced spin-density wave (SDW) fluctuations far above the bulk transition temperature[36]. However, a direct visualization of atomic-scale nematic pattern is still missing, especially for the most representative family of IBSCs, $Ae$Fe$_2$As$_2$ ($Ae$ = Ca, Sr, Ba). So far, only unidirectional electronic nanostructures around impurities, a characteristic feature of most IBSCs[37,38], have been ever observed in Co-doped CaFe$_2$As$_2$[39,40]. Recently, signatures of atomic-scale orbital order have been observed using specially prepared STM tips on the surface of FeTe$_{0.55}$Se$_{0.45}$[41], underscoring the significance of investigating atomic-scale nematicity on clean FeAs planes.

To complement this long-sought SI-STM investigation of atomic-scale nematicity in $Ae$Fe$_2$As$_2$, particularly on the essential FeAs plane, high-quality BaFe$_2$As$_2$ thin films were successfully grown on Nb-doped SrTiO$_3$(001) substrates using molecular beam epitaxy (MBE) (**Methods**). This enables unprecedented imaging of atomic-scale nematicity with SI-STM, modulated by 1/8 Ba adsorption and consequently characterized by a commensurate spatial periodicity of 4$a_{\text{Fe}}$, on the atomically flat FeAs planes. Intriguingly, the quantified electronic nematic parameters are spatially homogeneous within orthogonally oriented nematic domains and experience a sign change at approximately 30 meV. We find that the sign change and energy-dependent nematicity can be understood through the opposite orbital splitting of $d_{xz}/d_{yz}$ between electron and hole pockets, reflecting an unusual momentum and energy dependence of the orbital order in ferropnictides[22,23]. These atomic-scale electronic nematicity further manifests as unidirectional electronic nanostructures in the vicinity of single impurities and is notably suppressed by cobalt (Co) substitution for Fe atoms.

## Results

### Atomic-scale imaging of nematicity

Figure 1a illustrates a representative in-plane resistivity versus temperature curve for a 10 unit-cell (UC) BaFe$_2$As$_2$ film, revealing a pronounced decline near 134 K. This anomaly signifies a phase transition from the tetragonal paramagnetic state to the orthorhombic antiferromagnetic state at $T_{\text{N}}$, mirroring the behavior reported in its bulk counterpart[42]. Concurrently, an electronic nematic state sets in below $T_{\text{N}}$, simulating considerable interest within the research community of IBSCs. In order to facilitate direct imaging of this unusual nematic state with SI-STM, we prepared $BaFe_2As_2$ films on Nb (0.5-wt%)-doped $SrTiO_3$ substrates. A typical STM topography $T(\boldsymbol{r})$ of as-grown $BaFe_2As_2$ films is shown in Supplementary Fig. 1a, which often presents two categories of terminated surfaces. One surface is flatter, FeAs-terminated, and modulated by 1/8 Ba adsorption, resulting in an ordered $2\sqrt{2} \times 2\sqrt{2}$ reconstruction (Fig. 1b-d and Supplementary Fig. 1b), while the other is Ba-terminated, featuring a $2 \times 2$ surface reconstruction and a rough surface caused by randomly distributed adatoms (Supplementary Fig. 1c). On FeAs surfaces, the sparsely adsorbed Ba adatoms periodically occupy the hollow sites of the underlying Fe lattice, as schematically drawn in Fig. 1d. The identity of FeAs termination is confirmed by our visualization of the topmost As atoms surrounding the Ba adatoms, which manifest as rings at low sample biases (Fig. 1d and Supplementary Fig. 1b,d), and by the Co substitution for Fe (Supplementary Fig. 1e). Those Ba adatoms induce highly localized peaks in density of states (DOS) around 4.3 mV (see the top curve in Fig. 1e), occurring only within a radius of $\sim a_{Fe}$ from every Ba adatom (Supplementary Fig. 2). This enables us to directly access inherent properties of the FeAs plane by probing the differential conductance spectra $g(\boldsymbol{r}, V) \equiv dI/dV(\boldsymbol{r}, V)$ at hollow and bridge sites relative to the adsorbed Ba lattice. As illustrated in Fig. 1e, the representative $dI/dV$ spectra, particularly those acquired at the hollow sites, exhibit gap-like depletions in DOS near the Fermi level ($E_F$), similar to that previously identified in NaFeAs[43]. This gap most possibly originates from the spin density wave (SDW) and/or Fermi surface nesting-driven band hybridization between electron pockets around the M point and hole pockets at the $\Gamma$ point in the Brillouin zone (BZ)[22,44,45].

Upon closer inspection, one immediately notices a subtle yet significant difference in the $dI/dV$ spectra measured at two distinct bridge sites along the two orthogonal ($a$ and $b$) Fe-Fe directions, denoted as $B_x$ and $B_y$ (Fig. 1d), respectively. This difference is actually reflected in $T(\boldsymbol{r})$ images taken at low sample voltages. As illustrated in Fig. 1c, the Ba adatoms appear slightly elongated along one of the Fe-Fe directions, resulting in stripe-like patterns with a unique periodicity of $4a_{Fe}$. The stripes alternate in spatial orientation, as highlighted by the contrast-enhanced $T(\boldsymbol{r})$ image in Fig. 1f, giving rise to two orthogonally oriented domains. For convenience, we define the sites with Ba elongation along the $x$ direction as domain I and those along the $y$ direction as domain II for $T(\boldsymbol{r})$ images taken at $V \sim 20$ mV. These observations of atomic-scale rotational symmetry breaking underscore the inherent anisotropy, indicative of nematicity, within the underlying Fe plane. To the best of our knowledge, this constitutes the first direct observation of atomic-level nematicity in ferropnictides, previously unreported in earlier studies[36,40,41], made possible by the exposure of a clean FeAs plane. Unlike FeSe[39,46], the nematic domain wall (DW) observed here is exceptionally sharp and involve no structural distortion (Fig. 1c,f). This indicates an electronic origin for the observed nematicity, which can decouple from the structural anisotropy in BaFe$_2$As$_2$.

The observation of such pronounced atomic-level nematicity provides a unique opportunity to investigate its spatial dependence. We quantify the electronic nematicity by examining the difference of $T(r)$ values between the $B_x$ and $B_y$ sites within every $2\sqrt{2} \times 2\sqrt{2}$ superstructure of the adsorbed Ba adatoms. To achieve this, we first determine the spatial coordinates of all bridge sites surrounding each Ba adatom at $R_{ij}$ (Fig. 1g, top panel and Supplementary Fig. 3a), denoted as $R_{ij} \pm 2a_{Fe}\hat{x}$ along the $x$ axis and $R_{ij} \pm 2a_{Fe}\hat{y}$ along the $y$ axis, respectively. Evidently, the $T(r)$ values at $B_x$ and $B_y$ sites exhibit an apparent difference within a certain nematic domain, reversing between the distinct orthogonal domains (Fig. 1g, bottom panel). Therefore, a nematic order parameter $N_T(r)$ can be readily defined as the relative difference between mean $T(R_{ij} \pm 2a_{Fe}\hat{x})$ and $T(R_{ij} \pm 2a_{Fe}\hat{y})$ values around each Ba adatom, namely

$$N_T(r) = \frac{(T(R_{ij} + 2a_{Fe}\hat{x}) + T(R_{ij} - 2a_{Fe}\hat{x})) - (T(R_{ij} + 2a_{Fe}\hat{y}) + T(R_{ij} - 2a_{Fe}\hat{y}))}{T(R_{ij} + 2a_{Fe}\hat{x}) + T(R_{ij} - 2a_{Fe}\hat{x}) + T(R_{ij} + 2a_{Fe}\hat{y}) + T(R_{ij} - 2a_{Fe}\hat{y})}. \quad (1)$$

Figure 1h presents our measured $N_T(r)$ from the field of view (FOV) in Fig. 1c, where the continuous image is generated using biharmonic spline interpolation between Ba sites (Supplementary Fig. 3b). Aside from a few isolated impurities, the nematic order parameter $N_T(r)$ is spatially homogenous in magnitude, despite a sign change between the orthogonal domains. Remarkably, the nematic domain walls are narrower than $2a_{Fe}$ (Fig. 1h and Supplementary Fig. 3c) and diminish in $T(r)$ measurements at a high sample voltage of $V = 1.0$ eV (Fig. 1b and Supplementary Fig. 3d). With the disappearance of the domain walls, the statistical nematic order parameters in the $N_T(r)$ map become predominantly centered around zero (Supplementary Fig. 3e). This stands in stark contrast to a bimodal distribution of $N_T(r)$ in Fig. 1i, originating from the orthogonal nematic domains observed at $V = 20$ mV.

**Energy dependence of nematicity**

The fact that electronic nematicity becomes detectable from $T(r)$ only at low $V$ prompts further investigation into its energy dependence. This can be readily achieved by mapping the conductance $g(r, V)$ at varying energy $E = eV$ (where $e$ is the elementary charge). Figure 2a displays a larger FOV $T(r)$, in which a series of d$I$/d$V(r, V)$ maps were collected with $V$ varying from -20 meV to 20 meV. As depicted in Fig. 2b, the nematicity-driven $4a_{Fe}$ stripes and the orthogonal domains become more prominent in these low-energy d$I$/d$V$ maps. This is further supported by Fast Fourier-transform (FFT) analysis, which reveals unequal intensities between the $(\pm 0.25, 0)2\pi/a_{Fe}$ and $(0, \pm 0.25)2\pi/a_{Fe}$ peaks (circled in black) measured from the identical category of nematic domains (Fig. 2c). Notably, this intensity difference is also reversed between the orthogonal domains. Without loss of generality, the nematic order parameter can be similarly read from the energy-resolved $g(r, E = eV)$ maps as

$$N_g(r, E = eV) = \frac{\left(g(R_{ij} + 2a_{Fe}\hat{x}, V) + g(R_{ij} - 2a_{Fe}\hat{x}, V)\right) - \left(g(R_{ij} + 2a_{Fe}\hat{y}, V) + g(R_{ij} - 2a_{Fe}\hat{y}, V)\right)}{g(R_{ij} + 2a_{Fe}\hat{x}, V) + g(R_{ij} - 2a_{Fe}\hat{x}, V) + g(R_{ij} + 2a_{Fe}\hat{y}, V) + g(R_{ij} - 2a_{Fe}\hat{y}, V)}. \quad (2)$$

Figure 2d,e represent our measured $N_g(r, E)$ derived from the d$I$/d$V$ maps (i.e. Fig. 2b) and their statistical histograms across various energies, respectively. Once again, the nematic order parameter shows substantial spatial homogeneity within every domain, with a typical dimension of a few tens of nanometers. This length scale is comparable to those observed in epitaxial FeSe films[35,38,39], suggesting a common tendency toward electronic nematicity in ferropnictides and iron chalcogenides.

Notably, $N_g(r, E)$ exhibits prominent energy dependence, as reflected by the varying separation between local maxima in its energy-dependent bimodal distributions (Fig. 2e). Apparently, a larger separation between the local maxima implies a stronger electronic nematicity. For clarity, we denote the evolution of the two distinct maxima, contributed by I and II domains, with red and blue ribbons, respectively. One immediately notices a reduction in electronic nematicity near $E_F$, probably due to orbital hybridization between the hole and electron pockets[44], and a near-complete disappearance of the electronic nematicity around 20 meV. To shed additional light on these variations, we examine $N_g(r, E)$ within domain II across a broader energy range in Fig. 2f. Surprisingly, the $N_g(r, E)$ becomes positive and experiences a notable sign change around 30 meV, vice versa within domain I. The sign change in electronic nematicity has further been corroborated by directly comparing the site-specific d$I$/d$V(r, V)$ spectra at $B_x$ and $B_y$. By subtracting the spatially averaged $g(B_y, V)$ from $g(B_x, V)$, a sign change around 30 meV can be clearly revealed in the electronic DOS difference $D_g(E = eV) = g(B_x, V) - g(B_y, V)$, accompanied by a significant reduction in $D_g(E)$ near $E_F$ (Fig. 2g,h). Remarkably, the energy-resolved $D_g(E)$ and therefore the nematic order parameters are precisely reversed between domains I and II. It's worth noting that interference effects induced by an anisotropic STM tip can potentially give rise to an energy-dependent anisotropic DOS, which is highly sensitive to the tip's anisotropic parameter[47]. To evaluate this possibility, we have repeatedly replaced and reconditioned the STM tips throughout during our measurements and consistently observed reproducible results (Fig. 1 and Fig. 2). A definite conclusion would require systematic measurements within the same FOV under varying tip conditions – a method that is currently not feasible in our STM measurements. Nevertheless, as discussed later, the novel energy-dependent nematicity observed here shows good agreement with results from ARPES measurements and theoretical calculations, suggesting that the sign change near 30 meV is most likely an intrinsic property of the FeAs plane.

We emphasize that the electronic nematicity remains robust across various $BaFe_2As_2$ films and thickness. In a thinner $BaFe_2As_2$ film of 3 UC, which is somewhat influenced by epitaxial strain (Supplementary Fig. 4), the nematic order parameter $N_g(\boldsymbol{r}, E)$ exhibits a similar energy dependence, with a sign change near the similar energy of $\sim 30$ meV (Supplementary Fig. 5). At an elevated temperature of 78 K, well below $T_N$, the nematicity remains essentially unchanged, with a notable exception of the absence of reduction in $N_g(\boldsymbol{r}, E)$ near $E_F$ (Supplementary Fig. 5b,c). This can be reasonably understood as a result of reduced orbital hybridization between the hole and electron pockets, most probably linked to the suppression of antiferromagnetic order and thus reduced orbital hybridization at elevated temperatures.

## Impurity and Co doping effects on electronic nematicity

Next, we study the response of electronic nematicity to native impurities in as-grown $BaFe_2As_2$ films. As shown in Fig. 2a, two types of single impurities, appearing as bright protrusions and dark voids, decrease in number after post-annealing under ultrahigh vacuum conditions. We thus attribute these features as As adatoms and interstitials, which both have minimal influence on the electronic nematicity. In contrast, a third type of impurity, highlighted by a pink square in Fig. 2a, is registered at the Ba adatom sites and becomes more prevalent after a similar annealing process. We therefore ascribe the presence of this impurity to Ba desorption, considerably impacting the nearby electronic DOS and inducing unidirectional electronic nanostructures (Fig. 3a). These nanostructures are found to orient perpendicularly between domains I and II, as clearly shown in Fig. 3b,c. Nevertheless, the characteristic length scale of $\sim 8a_{Fe}$ of the symmetry-breaking nanostructures within both domains is comparable to that previously observed in the sister compound $Ca(Fe_{1-x}Co_x)_2As_2$, where Co atoms serve as the scattering centers[40,41]. Figure 3d,e depict the atomic-scale formation mechanism of the unidirectional electronic nanostructures by drawing the direction-dependent electronic DOS across different Ba adatom sites. First, the difference $D_g(E)$ in electronic DOS along the $x$ and $y$ directions is substantially suppressed at the Ba vacancy sites, resulting in a negligibly small $N_g(\boldsymbol{r}, E)$ there (Fig. 3b,c). Second, and most importantly, $D_g(E)$ is sign-reversed between two nearest Ba adatoms of the Ba vacancy along the $y$ direction within domain I, accompanied by an enhancement of $D_g(E)$ at the other nearest Ba adatoms along the $x$ direction, while the opposite behavior holds true within domain II. Consequently, the electronic DOS around Ba vacancies are significantly suppressed along one of the Fe-Fe directions (Fig. 3a-c), analogous to the observations in FeSe[38,39]. Note that the length scale of the unidirectional electronic nanostructures is approximately $16a_{Fe}$ in FeSe[39], which is apparently distinct from $8a_{Fe}$ observed in ferropnictides.

In addition to Ba vacancies, we investigate the influence of Co substitution for Fe atoms on electronic nematicity. Supplementary Fig. 6a shows a representative low-voltage $T(\boldsymbol{r})$ image taken from an optimally-doped $Ba(Fe_{1-x}Co_x)_2As_2$ ($x \sim 0.06$) thin film (10 UC), exhibiting the highest critical temperature $T_c \sim 30.5$ K (Supplementary Fig. 7). Using the same analysis method, we calculate $N_T(\boldsymbol{r})$ from this $T(\boldsymbol{r})$, which manifests as a maze-like pattern without discernible DWs in Supplementary Fig. 6b. This contrasts markedly with the uniform $N_T(\boldsymbol{r})$ and sharp nematic DWs identified in $BaFe_2As_2$ (Fig. 1h), indicating a significant suppression of the long-range nematic order by Co doping. Notably, the full-width at half maxima of the $N_T(\boldsymbol{r})$ histogram ($\sim 0.25$) is largely increased (Supplementary Fig. 6c), which is consistent with strong nematic fluctuations in optimally doped ferropnictides[48,49]. This sheds atomic-scale light on the intertwining of electronic nematicity with superconductivity in IBSCs. By substituting all Fe with Co atoms, we find a similar $2\sqrt{2} \times 2\sqrt{2}$ surface reconstruction on isostructural $BaCo_2As_2$ (Supplementary Fig. 6d). However, no electronic nematicity is revealed on the CoAs plane in Supplementary Fig. 6e, and the $N_T(\boldsymbol{r})$ values show a narrow distribution centered near zero (Supplementary Fig. 6f). This compellingly reveals that the observed electronic nematicity has a unique origin from the FeAs plane, rather than being influenced by any external factors.

## Discussion

Our experimental observations of this novel energy-resolved nematic order parameter, along with its significant sign change around 30 meV, offer crucial atomic-scale evidence of electronic nematicity in ferropnictides. The strong electron and/or spin correlation inherent in IBSCs lifts the degeneracy between $d_{xz}$ and $d_{yz}$ orbitals of Fe atoms around $T_N$[50-52]. This lifting leads to an inequivalent occupation of electrons within these orbitals, producing a significant difference in DOS around each Fe atom along the $x$ and $y$ directions. The disparity ultimately manifests as the atomic-scale electronic nematicity, as observed in our study (Fig. 2b). Moreover, the sign change of the nematic order parameter can be readily interpreted as a consequence of the contrasting splitting behavior of the Fe $d_{xz}/d_{yz}$ orbitals between the electron and hole pockets, as visualized by momentum-resolved ARPES measurements[22,23]. For the electron pockets near the M point of the BZ, the $d_{xz}$ orbital is energetically higher than the $d_{yz}$ orbital. This leads to a greater occupation of electrons in the $d_{yz}$ orbital, generating a negative $N_g(r, E)$. Conversely, the hole pocket centered at the $\Gamma$ point shows the opposite behavior, with more electrons occupying the $d_{xz}$ orbital, resulting in a positive $N_g(r, E)$. The crossover in dominant orbitals from electron to hole pockets takes place near 30 meV, aligning with the band top of the central hole pocket observed in ARPES[53,54], thereby highlighting the consistency between real-space and momentum-space measurements. Consequently, the sign of $N_g(r, E)$ changes from negative to positive within domain I, or from positive to negative within domain II where the $x$ and $y$ interchanges (Fig. 2), as we measure the energy-resolved $N_g(r, E)$ from the higher to lower energies. Alternatively, a ferro-orbital order has also been explored as a potential driving force of electronic nematicity in IBSCs, being lack of variations in momentum and energy[55]. This perspective stands in contrast to our findings, which underscores the unusual feature of electronic nematicity on the microscopic level. Combined with previous band structure measurements[22,23,53,54] and a wide range of experimental evidence[7,34], we reasonably infer that the nematicity observed is intimately linked to the orbital degrees of freedom, potentially coupled with anisotropic magnetic interaction, rather than to charge modulations as previously reported in IBSCs[34-40].

The orbital order within the Fe sublattice, characterized by inherent $a_{Fe}$-spaced nematic stripes and schematically shown in Fig. 4a, appears to be largely modulated by the presence of Ba adatoms (Fig. 4b). Specifically, the Ba adatoms strongly interact with the nearby electronic states within a radius of $a_{Fe}$, causing a local disruption of the orbital order at the four nearest-neighbor Fe atoms beneath the Ba adatoms (Supplementary Fig. 8), marked by the green balls in Fig. 4b. However, the other Fe atoms maintain their orbital order to minimize kinetic energy. This observation underscores the selective influence of Ba adatoms on the orbital order in the underlying Fe sublattice, enlarging the periodicity of nematicity-induced stripes from $a_{Fe}$ to $4a_{Fe}$. We highlight that the inequivalence of electronic DOS at $B_x$ and $B_y$, a fundamental cause of the observed electronic nematicity, is derived from the enhanced orbital overlaps along the lobe direction of the favored $d_{xz}$ or $d_{yz}$ orbital in the Fe plane (Fig. 4b). Consequently, the $4a_{Fe}$-spaced stripes are distinctly observed in $g(\boldsymbol{r}, E)$, oriented by $90^\circ$ not only between the orthogonally nematic domains I and II (Fig. 4c,d) but also as the energy $E$ is decreased across 30 meV (Supplementary Fig. 9). This straightforward phenomenological model effectively accounts for our experimental observations and underscores the atomic-scale electronic nematicity associated with orbital order.

Though microscopic electronic nematicity has been intensely studied in strongly correlated materials, atomic-scale imaging of this phenomenon remains limited, with most of the observations manifesting as maze-like patterns[35,36] and unidirectional electronic nanostructures in the vicinity of impurities[38-41]. Our direct visualization and systematic quantification of electronic nematicity at the atomic level, particularly on the FeAs plane, is unprecedented in ferropnictides. These findings, together with our elucidation of energy-resolved electronic nematicity and its sign reversal closely associated with the $d_{xz/yz}$ orbital order, provide essential insights toward constructing a microscopic model of electronic nematicity. Looking ahead, our research opens new avenues for studying the intricate interplay between electronic nematicity and superconductivity in ferropnictide films, which could lead to a deeper understanding of the underlying mechanism of unconventional superconductivity. Furthermore, investigating the impacts of varying doping levels, epitaxial strain, or external perturbations on the nematic landscape could yield valuable insights into the dynamics of these systems, potentially guiding the intentional design of unusual materials with tailored electronic properties and improved superconductivity.

**Methods**

**MBE growth.** Our Ba(Fe, Co)$_2$As$_2$ films were fabricated on TiO$_2$-terminated SrTiO$_3$ substrates under a temperature of ~ 520$^\circ$C and a vapor pressure of As exceeding $10^{-6}$ Torr. High-purity Ba (99.9%), Fe (99.995%) and Co (99.995%) metal sources were co-evaporated from standard Knudsen diffusion cells, with all flux calibrations performed *via* a Quartz Crystal Microbalance (QCM). The Co doping level was determined by calculating the its relative atom concentration in the total concentration of Co and Fe, while the film thickness is controlled by the deposition configuration and rate. Prior to MBE growth, the SrTiO$_3$ (001) substrates were annealed to ~ 1200$^\circ$C to obtain atomically flat $TiO_2$ surfaces. Afterward, the $BaFe_2As_2$ samples were annealed at about 550°C under ultra-high vacuum conditions to eliminate any residual As molecules from the surface.

**STM measurement.** All STM experiments were performed in a low-temperature STM facility (Unisoku. Co., Ltd.) under ultra-high vacuum conditions, with a base pressure below $2.0 \times 10^{-10}$ Torr. To enable precise STM functionality, BFCA films were prepared on 0.05-wt% Nb-doped $SrTiO_3(001)$ substrates. Polycrystalline PtIr tips, calibrated on MBE-grown Ag/Si(111) films, were used throughout the measurements at 4.5 K, unless otherwise specified. All STM topographic images were acquired in a constant current mode, while the tunneling d$I$/d$V$ spectra and maps were collected using a standard lock-in technique with a small a.c. modulation voltage ($V_{mod}$) at a frequency of $f = 983$ Hz. Here, $V_{mod}$ is typically set to 1/50 of the setpoint bias to ensure adequate energy resolution.

**Transport measurements**: After *in-situ* STM measurements, the sample were taken out of the ultra-high vacuum chamber for electrical measurement. The electrical resistivity was measured via a four-terminal configuration ($I = 1$ $\mu$A) in a commercial physical property measurement system (PPMS).

**Data availability**

Source data are provided upon request from the corresponding author.

**Acknowledgments**

This work was financially supported by the National Natural Science Foundation of China (Grant No. 12141403, Grant No.12134008, Grant No. 12474130) and the National Key R&D Program of China (Grant No. 2022YFA1403100).

**Author contributions**

C.L.S., X.C.M. and Q.K.X. conceived the project. Q.J.C., Y.W.W., M.Q.R. and C.C.L. synthesized the samples and performed the STM experiments. Z.X.D. carried out the transport measurements. All authors contributed to the data analysis and the manuscript.

**Additional information**

Supplementary information includes 9 figures and the corresponding figure captions.

Correspondence and requests for materials should be addressed to C.L.S., X.C.M., Q.K.X.

**Competing financial interests**

The authors declare no competing financial interest.

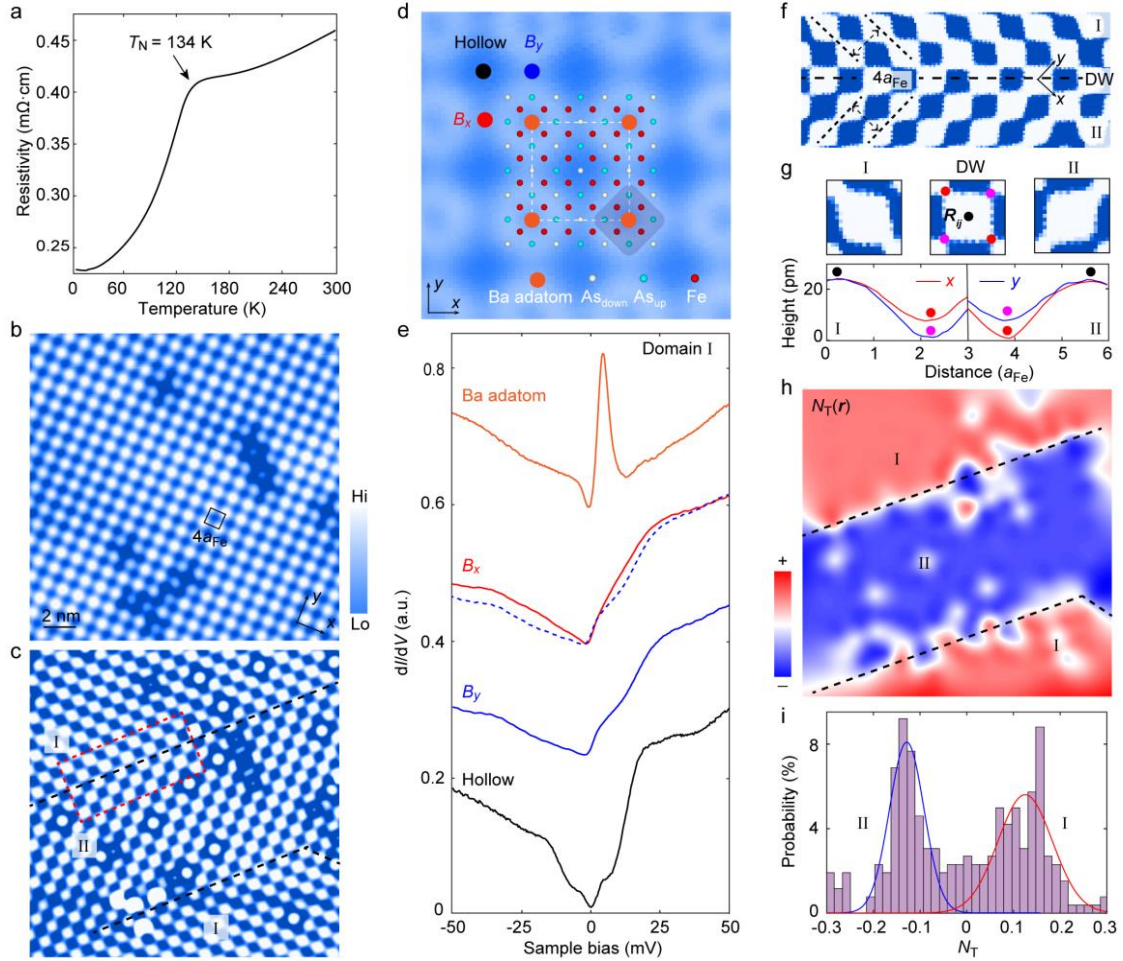

**Fig. 1 Atomic-scale nematicity on FeAs. a** Electrical resistivity versus temperature curve for a 10-UC BaFe$_2$As$_2$ film epitaxially grown on insulating SrTiO$_3$(001) substrate. The black arrow indicates the nematic transition temperature ($T_N$) around 134 K. **b** STM topography (20 nm × 20 nm) acquired on FeAs-terminated surface, exhibiting a 2 √2 × 2 √2 reconstruction induced by 1/8 Ba adsorption. The black square outlines the super unit cell formed by Ba adatoms. Setpoint: $V = 1.0$ V and $I = 50$ pA. **c** STM topography in the same FOV as in **b**, but taken at a different setpoint of $V = 20$ mV and $I = 100$ pA. The dashed black lines mark DWs separating orthogonally nematic domains I and II. **d** Structural model of 2 √2 × 2 √2 surface reconstruction due to Ba adsorption on the FeAs plane. Ring-like structure in the zoom-in STM topography (3.6 nm × 3.6 nm, $V = 100$ mV, $I = 100$ pA) is formed by four top-layer As atoms, with Ba adatoms adsorbed at their centers. **e** Site-resolved d$I$/d$V$ spectra measured within domain I, as indicated in **d**. The dashed blue line represents a vertical offset from the solid blue line to facilitate comparison between the spectra taken at the $B_x$ and $B_y$ sites. **f** Contrast-enhanced STM topography from the red-boxed region in **c**, highlighting dragged Ba adatoms and 4$a_{Fe}$-periodic stripes aligned along the Fe-Fe directions. **g** Zoom-in of Ba adatoms

within domain I, domain II and DW (top panel) as well as the line profiles along $x$ and $y$ directions within domains I and II (bottom panel), respectively. Colored dots mark Ba adatoms (black), $B_x$ (red) and $B_y$ (magenta) bridge sites. **h** Spatial map of nematic order parameter $N_T(\boldsymbol{r})$ extracted from STM topography $T(\boldsymbol{r})$ in **c**. **i** Histogram of $N_T(\boldsymbol{r})$ values from the nematic map in **h**, displaying a bimodal distribution.

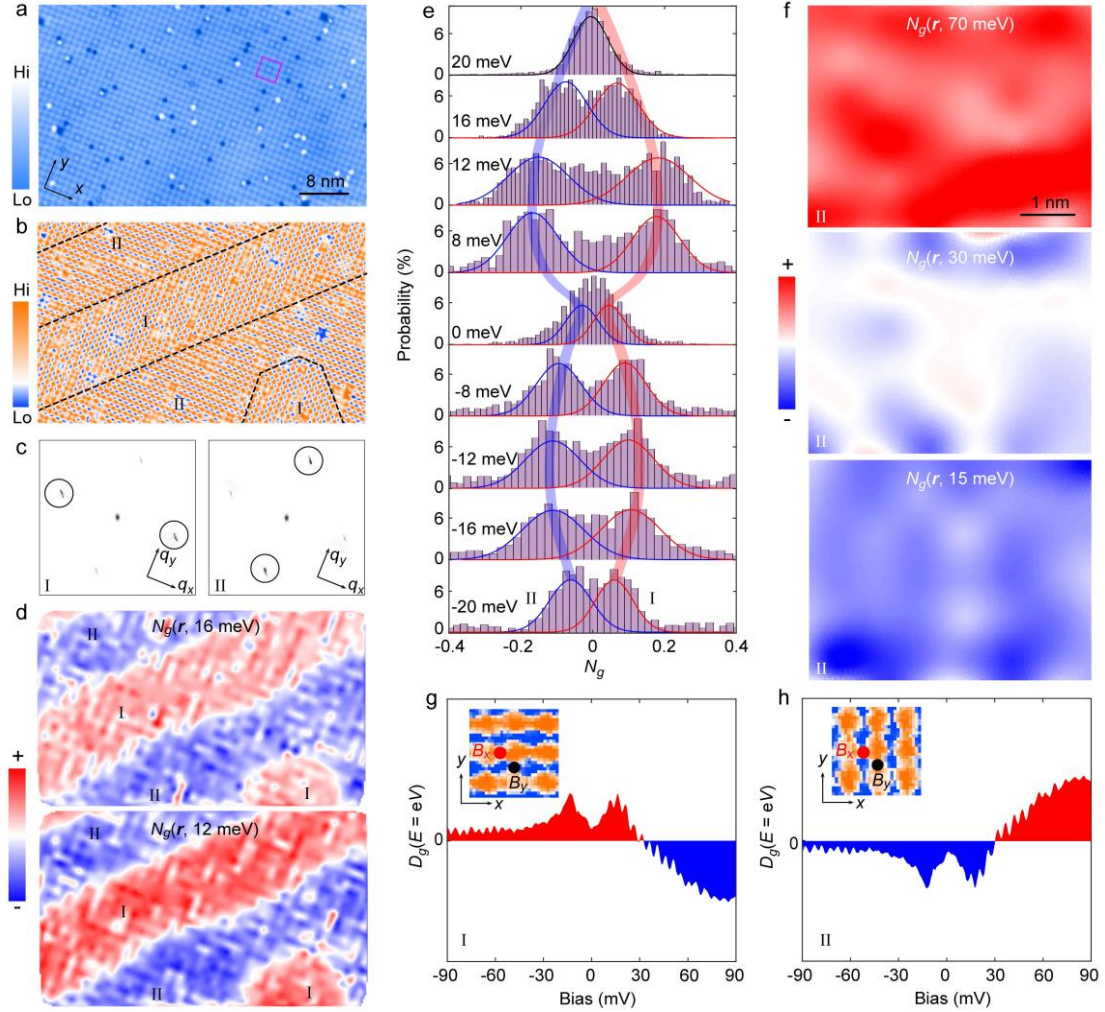

**Fig. 2 Energy dependence of nematic order parameter. a** Atomic-resolved STM topography ($V$ = 20 mV, $I$ = 500 pA) of the FeAs plane over a large FOV (55 nm × 34 nm). **b** d$I$/d$V$ map recorded at 12 meV from the FOV in **a**. Dashed lines indicate DWs separating homogeneous nematic domains. **c** Fourier-transformed images of domains I (left) and II (right) from the d$I$/d$V$ map in **b**, exhibiting inequivalent FFT peak intensities at ($\pm$ 0.25, 0)$2\pi/a_{Fe}$ and (0, $\pm$ 0.25)$2\pi/a_{Fe}$ as circled. **d** $N_g(\boldsymbol{r}, E)$ of the identical FOV as in **b**, measured at $E$ = 16 meV and 12 meV. **e** Histograms of $N_g(\boldsymbol{r})$ at various energies. The red and blue Gaussian fits correspond to the nematic order parameters within domains I and II, respectively. **f** Energy-dependent $N_g(\boldsymbol{r}, E)$ maps, revealing a sign change of the nematic

order parameter around 30 meV within domain II. **g,h** Spatially averaged differences in DOS ($D_g(\boldsymbol{r})$) between the $B_x$ and $B_y$ sites within domains I and II, respectively. Insets show zoom-in $g(\boldsymbol{r}, 12 \text{ meV})$ within domains I and II, with $B_x$ and $B_y$ sites explicitly marked.

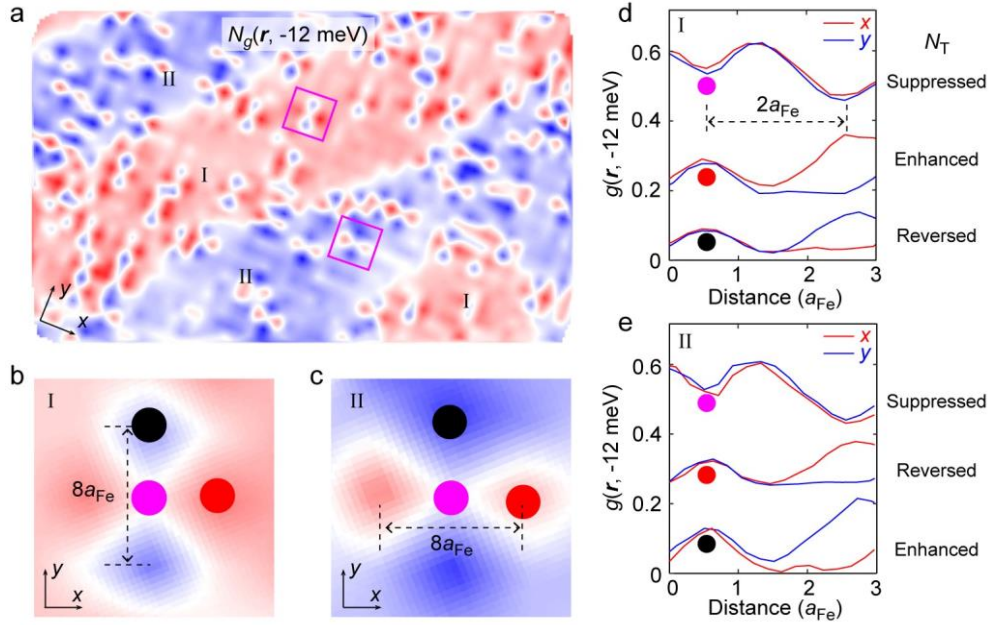

**Fig. 3 Unidirectional nematic pattern induced by single impurities. a** Nematic map $N_g(\boldsymbol{r})$ of the same FOV as in Fig. 2**d**, measured at -12 meV. Magenta squares mark Ba vacancy sites. **b,c** Enlarged $N_g(\boldsymbol{r})$ centered around Ba vacancies within domains I and II, inducing unidirectional nanostructures with a characteristic length scale of $8a_{\text{Fe}}$. **d** Line profiles across the pink, red and black circles within domain I, taken along the $x$ and $y$ directions. The nematic order parameter is significantly suppressed at Ba vacancy sites (pink circles), but enhanced and sign-reversed for the nearest Ba adatoms (red and black circles) along the $x$ and $y$ directions, thereby resulting in local symmetry breaking. **e** Same line profiles to **d**, but within domain II, showing a reversed nematicity.

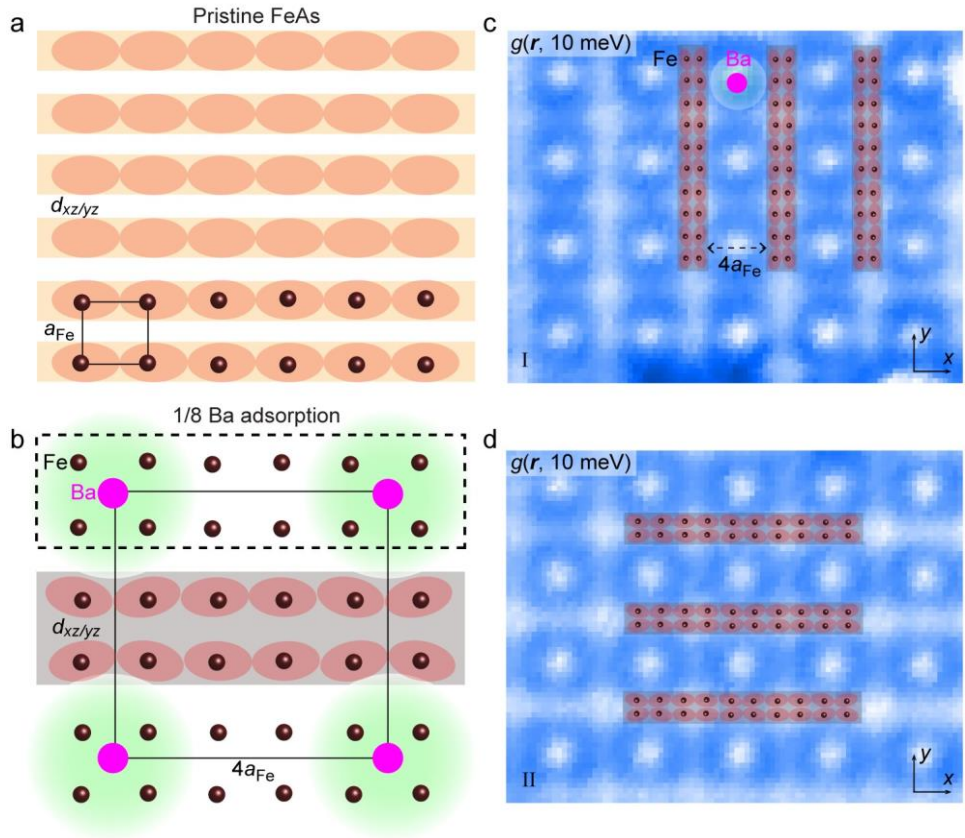

**Fig. 4 Phenomenological model of $4a_{Fe}$ stripes. a** Schematic illustration of the $d_{xz/yz}$ orbital order with a periodicity of $a_{Fe}$ in the pristine FeAs plane. **b** Schematic of the $4a_{Fe}$ stripes induced by 1/8 Ba adsorption. The local potential from every Ba adatom (indicated by green circles) suppresses the nematic order of the underlying Fe atoms within a radius of $\sim a_{Fe}$. This disrupts the nematic stripes within the dashed box, while the stripes within the filled gray box remain unchanged, resulting in fourfold periodic stripe as observed. **c,d** Zoom-in $g(\mathbf{r}, 10 \text{ meV})$ maps (5 nm × 4 nm) acquired within domains I and II, respectively, showing $4a_{Fe}$-periodic stripes that rotates by 90° between domains I and II.

**Atomic-scale imaging of electronic nematicity in ferropnictides**

Qiang-Jun Cheng[1], Yong-Wei Wang[1], Ming-Qiang Ren[1,2], Ze-Xian Deng[1], Cong-Cong Lou[1],

Xu-Cun Ma[1,3*], Qi-Kun Xue[1,2,3,4*], Can-Li Song[1,3*]

[1]*State Key Laboratory of Low-Dimensional Quantum Physics, Department of Physics, Tsinghua University, Beijing 100084, China*

[2]*Shenzhen Institute for Quantum Science and Engineering and Department of Physics, Southern University of Science and Technology, Shenzhen 518055, China*

[3]*Frontier Science Center for Quantum Information, Beijing 100084, China*

[4]*Beijing Academy of Quantum Information Sciences, Beijing 100193, China*

*Correspondence to: clsong07@mail.tsinghua.edu.cn, liuqh@sustech.edu.cn, xucunma@mail.tsinghua.edu.cn, qkxue@mail.tsinghua.edu.cn

This PDF file includes:

Supplementary Fig. 1 to Fig. 8

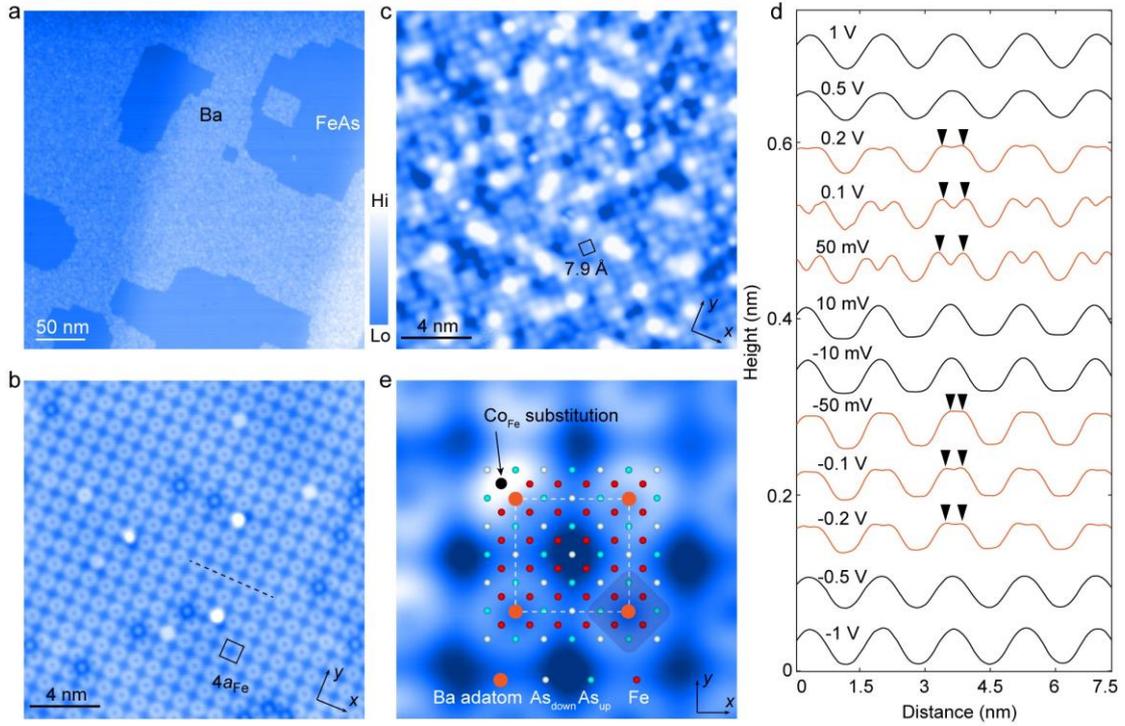

**Supplementary Fig. 1 STM characterization of as-grown BaFe₂As₂ films. a** STM topography in a representative BaFe₂As₂ thin film grown on a 0.5% Nb-doped SrTiO₃(001) substrate, exhibiting a coexistence of Ba- and FeAs-terminated surfaces, as labeled. **b** STM topography (20 nm × 20 nm, $V$ = 0.1 V and $I$ = 100 pA) showing ring-like structure formed by the top-layer As atoms. **c** Typical STM topography (20 nm × 20 nm, $V$ = 100 mV, $I$ = 50 pA) on Ba-terminated surface exhibiting a 2 × 2 reconstruction, outlined by a black square. Adsorbed atoms, visible as bright spots, are always observed across the surface. **d** Line profiles measured along the dashed line in **b** as a function of $V$, revealing the splitting at Ba adatom sites primarily occurring at low sample voltages (indicated by the arrow-marked double peaks). **e** STM topography (3.6 nm × 3.6 nm, $V$ = 100 mV, $I$ = 50 pA) of a Fe₀.₉₄Co₀.₀₆As plane overlaid with the atomic structural model proposed. A Co substitution for Fe (black circle) induces a bright petal on the ring.

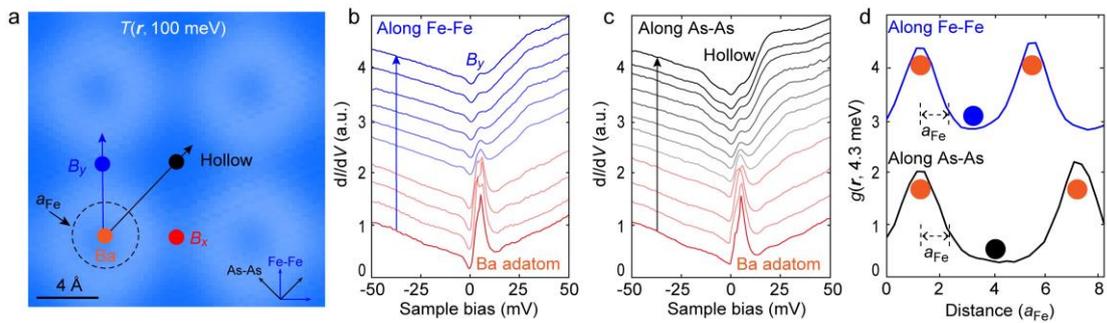

**Supplementary Fig. 2 Locality of DOS modulations induced by adsorbed Ba atoms. a** Atomic-resolved $T(r)$ (2 nm × 2 nm, $V$ = 100 mV, $I$ = 100 pA) on FeAs surface, highlighting the locations of adsorbed Ba adatom (orange circle), hollow (black circle) and bridge sites (blue and red circles)

relative to the adsorbed Ba lattice, respectively. **b,c** Line-cut d$I$/d$V$ spectra acquired along the Fe-Fe and As-As directions (along the trajectories indicated in **a**), showing a rapid decay of the Ba adatom-induced DOS peak at about 4.3 mV. **d** Spatial evolution of $g(\boldsymbol{r}, 4.3$ meV) measured along the Fe-Fe and As-As directions, presenting a characteristic radius of about $a_{\text{Fe}}$, beyond which the influence of adsorbed Ba diminishes significantly.

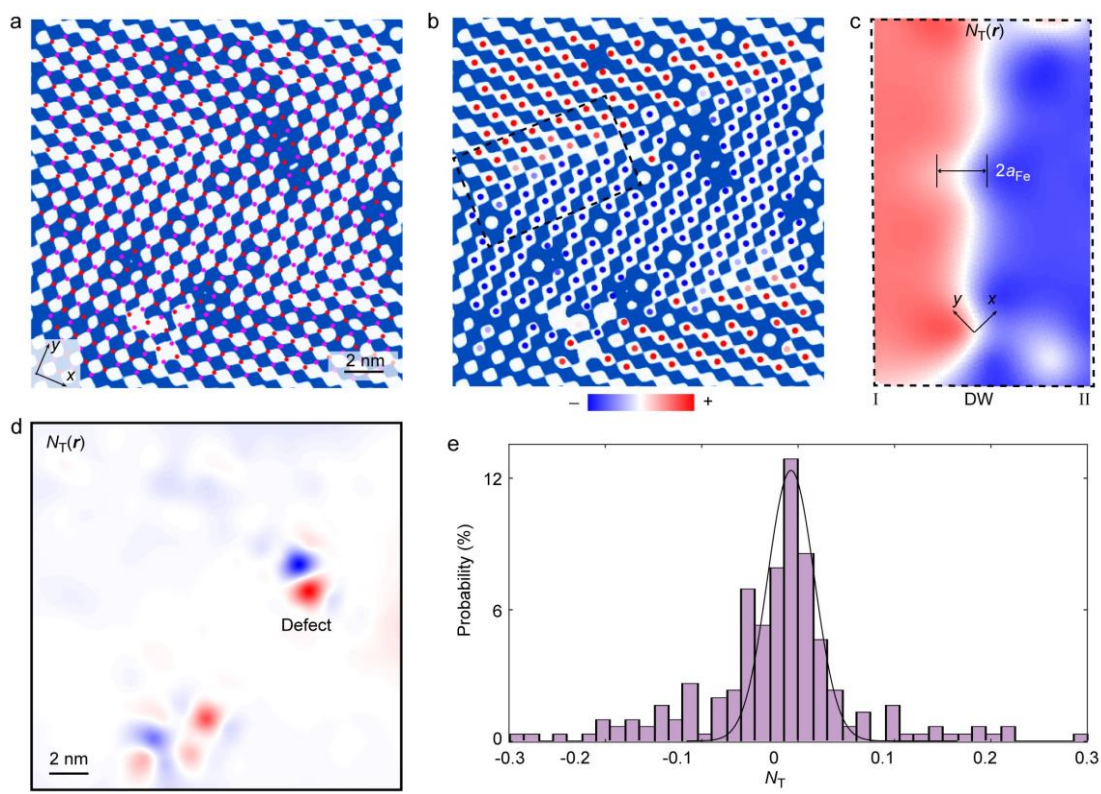

**Supplementary Fig. 3 Quantification of the nematic order parameter in real space. a** Atomic-resolved $T(\boldsymbol{r})$ (20 nm × 20 nm, $V = 20$ mV, $I = 50$ pA) on FeAs plane, with red/pink markers depicting the two distinct bridge sites ($B_x$ and $B_y$) between neighboring Ba adatoms. **b** Same topography as in **a**, overlaid with colored markers representing the onsite nematicity, defined by the relative height difference between the $B_x$ and $B_y$ bridge sites. **c** Nematic map, $N_{\text{T}}(\boldsymbol{r})$, within the dashed black box in **b**, obtained via a biharmonic spline interpolation of the onsite nematicity. The map reveals a sharp transition of the sign change of $N_{\text{T}}(\boldsymbol{r})$ across the DW, which is narrower than $2a_{\text{Fe}}$. **d** $N_{\text{T}}(\boldsymbol{r})$ measured from STM topography $T(\boldsymbol{r})$ measured at $V = 1.0$ V, showing no anisotropy across the entire FOV in **a**, except for the isolated defects. **e** Histogram of $N_{\text{T}}(\boldsymbol{r})$ in **d**. The black curve corresponds to the best fit of the experimental data to a single Gaussian function centered at zero, indicating the absence of nematicity at high energies.

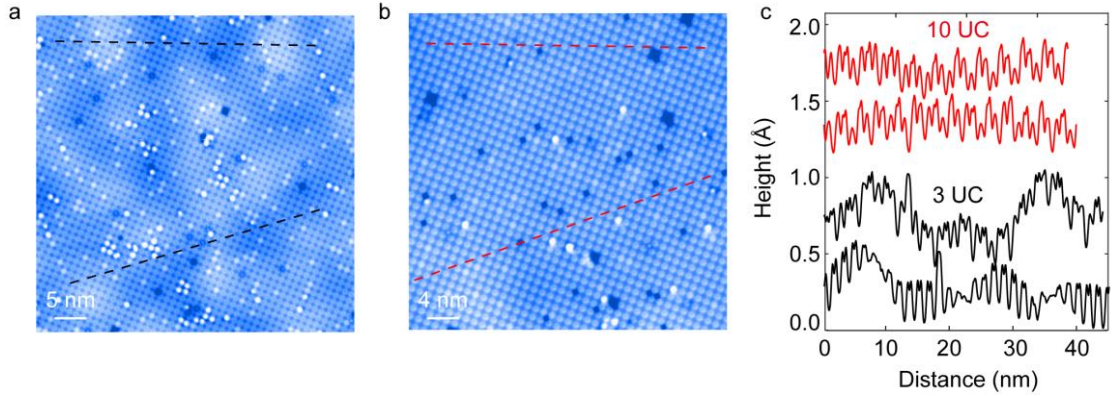

**Supplementary Fig. 4 Epitaxial strain in 3 UC BaFe₂As₂ film. a** Atomic-resolved $T(\boldsymbol{r})$ (50 nm × 50 nm, $V$ = 100 mV, $I$ = 100 pA) on FeAs plane of 3 UC BaFe₂As₂ thin film, where bright and dark regions indicate significant epitaxial strain. **b** Atomic-resolved $T(\boldsymbol{r})$ (40 nm × 40 nm, $V$ = 100 mV, $I$ = 100 pA) on FeAs plane of 10 UC BaFe₂As₂ thin film, with obviously weak epitaxial strain. **c** Line profiles taken along black and red lines, respectively. The large ups and downs superimposed on the lattice periodicity indicates the presence of significant strain in thinner sample (black lines).

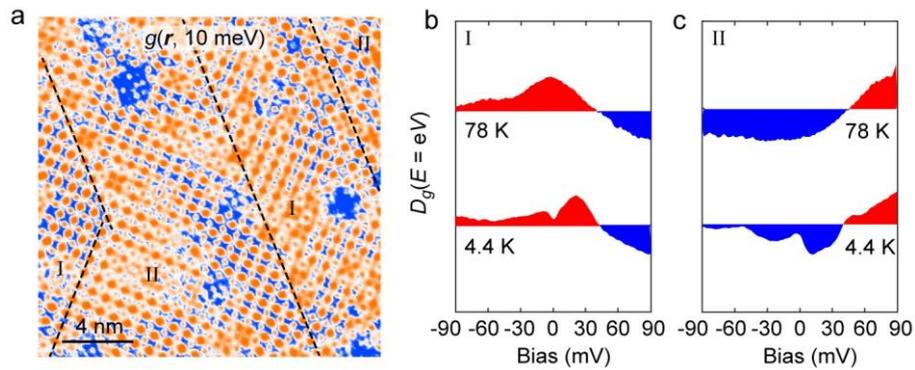

**Supplementary Fig. 5 Electronic nematicity in 3 UC BaFe₂As₂ film. a** $g(\boldsymbol{r}$, 10 meV) map (30 nm × 30 nm) taken in a representative 3 UC BaFe₂As₂ thin film, demonstrating the robust nematicity at 4.4 K. **b** Electronic DOS difference $D_g(E)$ between the $B_x$ and $B_y$ sites within domain I, measured at 4.4 K and 78 K. **c** $D_g(E)$ measured within domain II. While the suppression near $E_F$ is missing, the sign change in $D_g(E)$ occurs robustly at ~ 30 meV, irrespective of the measurement temperatures.

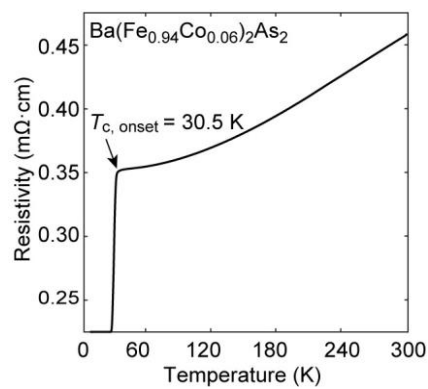

**Supplementary Fig. 6 Electrical resistivity versus temperature curve for a optimally-doped Ba(Fe$_{1-x}$Co$_x$)$_2$As$_2$ ($x$=0.06) thin film.** The film is epitaxially grown on insulating SrTiO$_3$(001) substrate. The black arrow indicates the onset critical temperature of superconductivity ($T_{c,\ onset}$) around 30.5 K.

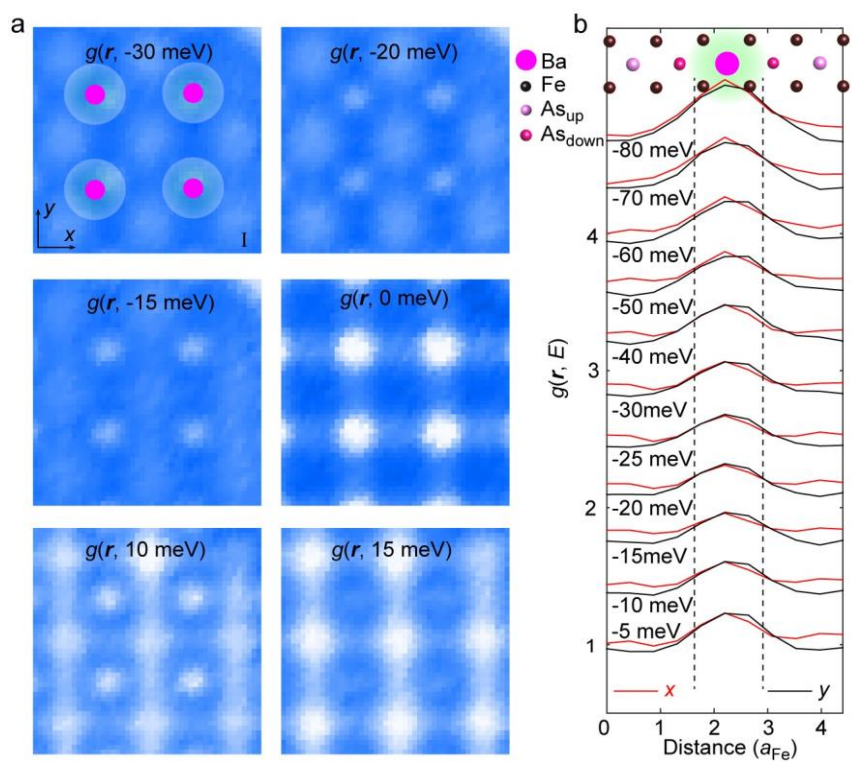

**Supplementary Fig. 7 Local suppression of electronic nematicity by the Ba adatoms. a** Energy-dependent $g(\boldsymbol{r}, E)$ maps (2.5 nm × 2.5 nm), taken on the FeAs surface within domain I. Ba adatom positions are marked by pink circles, and their influenced regions on the nearby DOS are highlighted by green circles. Within the green circles, the electronic DOS shows perfect $C_4$ rotational symmetry, while anisotropic DOS emerges between neighboring green circle. **b** Line profiles of $g(\boldsymbol{r}, E)$ along two orthogonal directions, showing consistent behavior within a radius of $a_{Fe}$ at all energies. Beyond $a_{Fe}$, the anisotropy of $g(\boldsymbol{r}, E)$ along the $x$ and $y$ axes becomes apparent, indicative of the emergence of electronic nematicity.

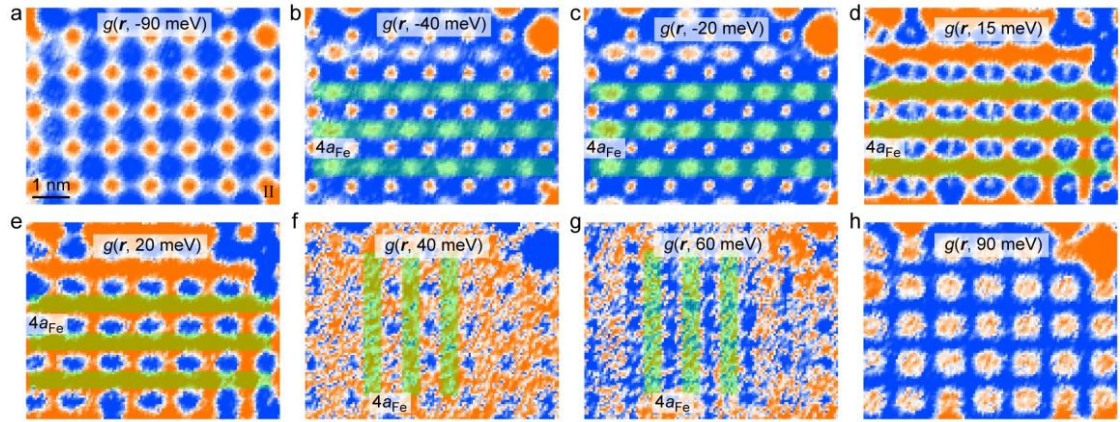

**Supplementary Fig. 8 Energy-dependent electronic nemeticity. a–h** Energy-dependent $g(\boldsymbol{r}, E)$ maps within domain II, collected over a 7.0 nm × 5.5 nm FOV. The $4a_{\mathrm{Fe}}$ stripe pattern exhibits a 90º rotation across a specific energy between 20 meV and 40 meV, reflecting an energy dependence of the electronic nematicity. Notably, the $C_4$ rotational symmetry is preserved at ± 90 meV.